# Silicon Metasurface Embedded Fabry-Pérot Cavity Enables High Quality Transmission Structural Color

**YUNXUAN WEI,[1,2] MING ZHAO,[1,*] ZHENYU YANG[1]**

[1]*School of Optical and Electronic Information, Wuhan National Laboratory for Optoelectronics, Huazhong University of Science and Technology (HUST), Wuhan 430074, China*

[2]*Ming Hsieh Department of Electrical and Computer Engineering, University of Southern California, Los Angeles, California 90089, USA*

**Corresponding author: zhaoming@hust.edu.cn*

**While nanoscale color generations have been studied for years, the high performance transmission structural color, simultaneously equipped with large gamut, high resolution, and optical multiplexing abilities, still remains as a hanging issue. Here, a silicon metasurface embedded Fabry-Pérot cavity is demonstrated to address this problem. By changing the planar geometries of metasurface units, the cavities provide transmission colors with 194% sRGB gamut coverage and 141111 DPI resolution, along with more than 300% enhanced angular tolerance. Such high density allows two-dimensional color mixing at diffraction limit scale. Beneficial from the polarization manipulation capacity of the metasurface, arbitrary color arrangements between cyan and red for two orthogonal linear polarizations are also realized. Our proposed cavities can be used in filters, printings, optical storages and many other applications in need of high quality and density colors.**

**Introduction.** Advanced structural colors, supposed to combine pure colors with ultrahigh spatial resolution, optical multiplexing abilities and low loss, are essential for the development of various optical devices and applications such as displays, CMOS sensors, information storage and encryption. Recent years, because of the potential to satisfy these desires, artificial nanostructures for both reflective and transmissive colors have been widely studied. Such structural colors are always generated by various resonances supported in nanoscale resonators [1, 2], including Mie resonances in dielectric scatters [3-6], surface plasmonic resonances excited in metallic structures [7-10], and Fabry-Pérot (FP) cavities [11-18]. Although having been achieved for reflective colors [6], simultaneously realizing extremely wide color gamut and ultrahigh resolution is still challenging for transmissive colors [19].

Fabry-Pérot (FP) cavity is one of the most widely applied approaches to generate transmissive structural colors [11-16, 20]. Traditional FP cavities need to change the cavity lengths to adjust the colors, which require complicated nanofabrication processes for integration [2]. Metasurfaces, composed of subwavelength unit arrays, have made planar FP cavities possible [13-15]. Previous works have explored their possibilities and developed related fabrication skills for both cavities with metallic mirrors and Bragg reflectors.

In this letter, we embed silicon (Si) metasurfaces into the FP cavities (meta-FP cavities) to generate transmissive structural colors with ultrawide color gamut, ultrahigh resolution and optical multiplexing abilities. Taking advantage of the small units, we demonstrate high density color mixing and build printings with varying hue and saturation based on the mixing pixels. To overall estimate the performance, the influence of the spatial resolution and the illumination angle are carefully discussed. Finally, polarization multiplexed colorful display is also enabled by cross shaped nanopillars.

**Results.** The metasurface unit [Fig. 1(a)], composed of Si [21] nano-cuboid with height $h_1$ and edge length $a$, is centrally embedded in the UV PDMS [22] cuboid lattice with height $h_2$ and period $P$. Si is selected to comprise the metasurface because of not only its high refractive index but also its high loss in short wavelengths. For the unit with $h_1$ = 50 nm, $h_2$ = 90 nm and $P$ = 180 nm, its phase and absorbance versus $a$ from 40 nm to 140 nm are simulated (Supplement 1) and shown in Fig. 1(b, c). With such units, short wavelength colors are mostly generated at small $a$ where transmittances are high, and partially absorbed at large $a$ to provide high purity long wavelength colors. Then as illustrated in Fig. 1(d), two Ag [23] films, with thickness $t$ = 35 nm, are separately added to the bottom and the top of the metasurface unit, to constitute the meta-FP unit upon a silica (SiO2 [21]) basement. The small cavity length of 90 nm ensures only the first order FP mode is excited in the visible range. Meantime, the gap $g$ between the nano-cuboid and the Ag film reduces the plasmonic coupling, helping maintain the mode purity. Fig. 1(e) shows the simulated relationship between the transmission spectrum and the edge length $a$. Insert pictures in Fig. 1(e) depict normalized electric field amplitude distributions in the $xz$ cross section plane, at the two markers respectively on plasmonic resonance and on FP resonance. The deep subwavelength period confines the plasmonic coupling peak to wavelengths around 400 nm, therefore reduces its impact on color purity. This fixed period also benefits printings with lattice matching between all units, leading to the tight integration of colorful pixels. The full width of half maximum (FWHM) of the FP transmission peak, with an average value of 20.5 nm, appears to reach the maximum of 26.2 nm at the wavelength of 434 nm and gradually reduces to 18.5 nm as resonant wavelength increases. The transmission spectrums are translated into color coordinates [Fig. 1(f)] using CIE 1931 chromaticity functions (Supplement 2). Calculation shows the curve linking color coordinates of edge length $a$ from 40 nm to 137 nm covers the largest area, revealing an ultrahigh gamut coverage of 194% sRGB, 143% Adobe RGB, and even 102% Rec. 2020. The coverage area also intersects 100% of sRGB, 100% of Adobe RGB, and 91% of Rec. 2020, meaning the generated colors are highly compatible with existing color systems.

Although in use of the lossy Si, the transmittance still reaches the maximum of 59% and an average of 43%. Fig. 1(g) depicts the absorption spectrum in Si. In short wavelengths, the nano-cuboid strongly enhances the absorption of the plasmonic induced transmissive mode and the background light, which improves the purity of the color from chartreuse to red. To evaluate the loss contribution at the FP peaks, we list the absorbances at the peak wavelengths, respectively in Ag films and Si, in Fig. 1(h). For small $a$, the light confined in the nano-cuboid raises as $a$ enlarges, resulting in increased absorption. When $a$ is large, due to the decreasing absorption of Si in long wavelengths, the absorbance drops. Though Si surely introduces loss, the absorbance of the nano-cuboid is still below the loss in Ag films.

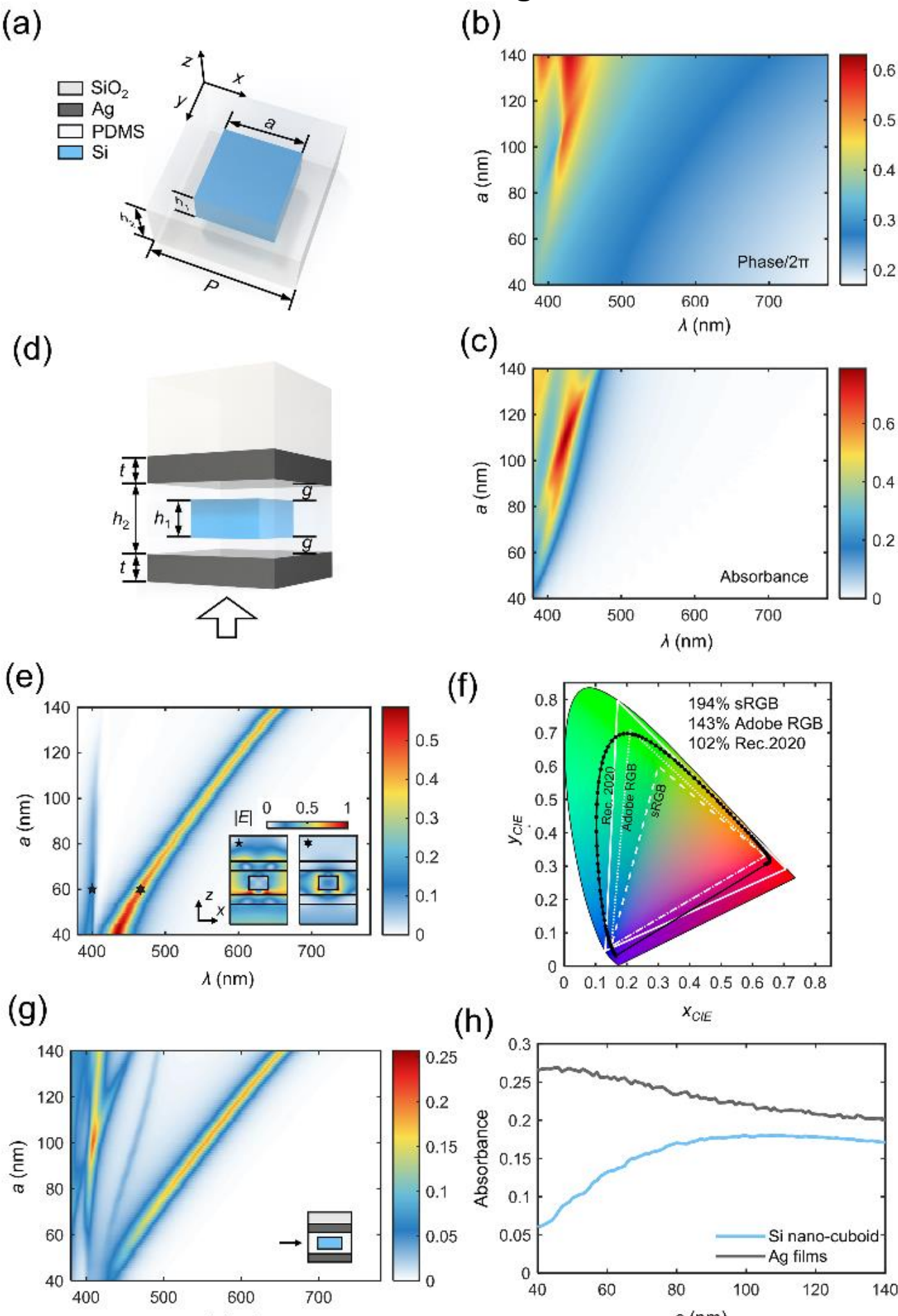

**Fig. 1.** Illustration of the Si meta-FP cavity and its properties. (a) Scheme of a metasurface unit. (b, c) Simulated phase and absorbance of the metasurface unit with varying edge length $a$, for $x$ polarized normal injection. (d) Scheme of a meta-FP unit. (e) Simulated transmission spectrum of the meta-FP unit, under $x$ polarized normal incidence imprinting from the bottom. Inserts show normalized electric field amplitude profiles in $xz$ plane cross sections, respectively at the pentagram marker and the hexagram marker. (f) Corresponding color coordinates plotted with black points. The calculated color gamut is circled with the black line. (g) The absorption spectrum of the embedded Si nano-cuboid. (h) Absorbance in the metasurface unit and the two Ag films, at the FP peak wavelength for each $a$.

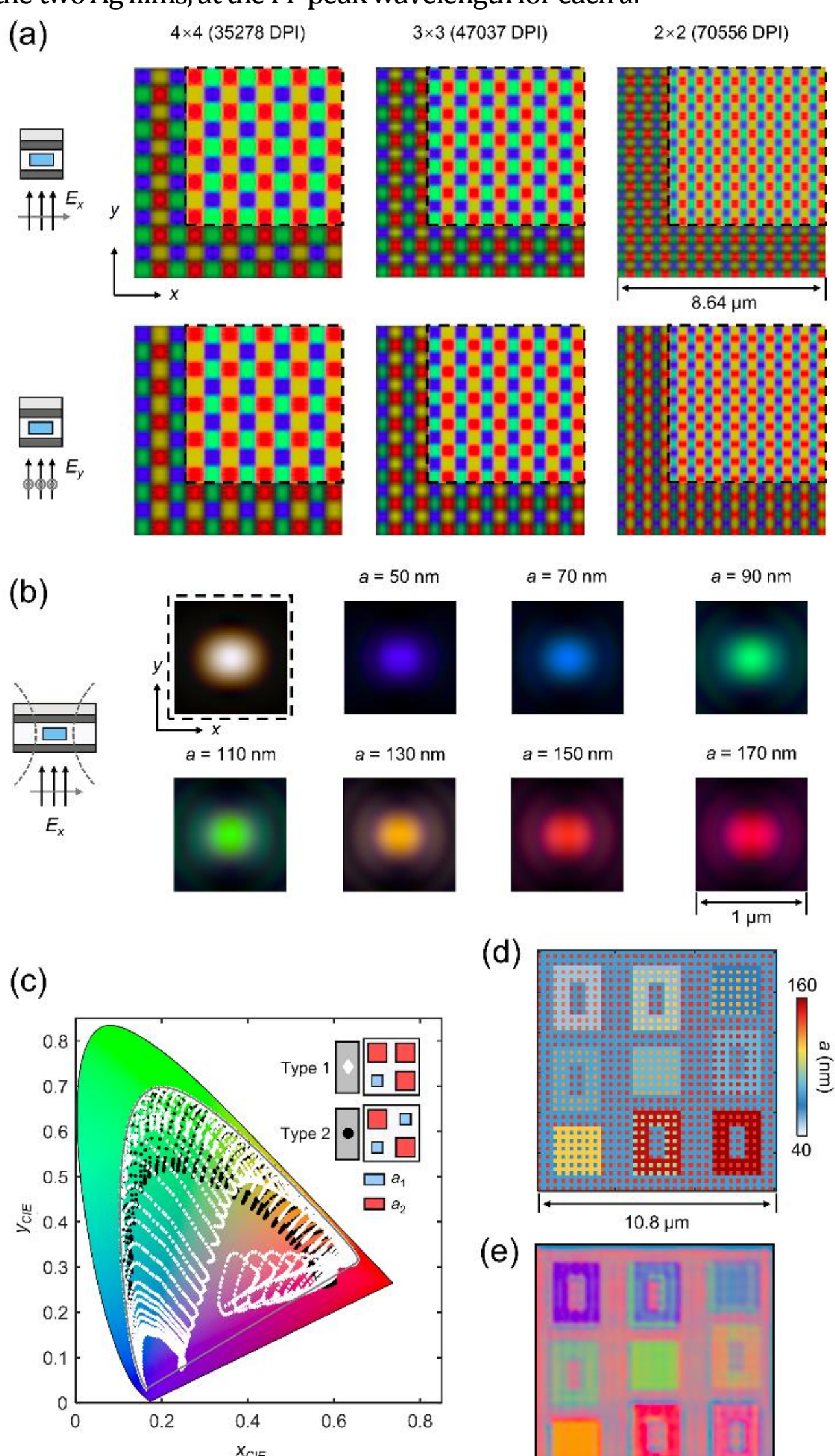

**Fig. 2.** Spatial resolution and printing tests for the meta-FP cavities. (a)Simulated periodic RGBY filter arrays, with each color pixel containing 4×4, 3×3, 2×2 meta-FP units. The filter array is illuminated by normally impinging light, under $x$ (upper) or $y$ (lower) linear polarization states. Brightness information in the dashed box is removed. (b) Transmissive color profiles of the same Gaussian beam passing through different FP cavities embedded single isolate nano-cuboids. Image in dashed box shows the color of the Gaussian beam going through the same space whereas without Ag films and the nano-cuboid. (c) Scheme of two types of mixing pixels and their simulated color coordinates, with edge length $a_1$ and $a_2$ varying from 40 nm to 160 nm. Blue and red squares represent two groups of units. Nano-cuboids in the two groups are with different edge lengths, of $a_1$ for blue and of $a_2$ for red. Gray line shows the 194% sRGB gamut of the meta-FP cavity. (d) Map of edge length $a$, for a printing made of a 30×30 Type

1 mixing pixel array. (e) Corresponding simulated color profile, without brightness information.

Then we characterize the resolution with periodic red-green-blue-yellow (RGBY) filter arrays. Each color corresponds to $a$ = 140 nm, 85 nm, 50 nm and 110 nm. The sizes of the color pixels are set to 4×4, 3×3, 2×2, corresponding to the spatial resolution of 35278 DPI, 47037 DPI, 70556 DPI. The colorful images are calculated from the simulated light fields at the $xy$ plane 15 nm above the top mirror (Supplement 3). Fig. 2(a) collects the colors from the three filter arrays, in which the brightness information in the dashed box is removed to better compare the hue. Obviously, all of them can generate highly pure and distinctive colors, with slight hue variations between two polarizations. As for the color pixel with 1×1 meta-FP unit, its minute size falls below the resolution limit ($0.61\lambda/n_{SiO2}$) for wavelengths larger than 433 nm, therefore is indistinguishable. Alternatively, we test the light transmitting the FP cavity embedded only one isolate nano-cuboid, using a Gaussian beam. Dashed box in Fig. 2(b) shows the transmission color when without Ag films and the nano-cuboid. Color distributions of the same beam passing through the complete meta-FP cavities are also listed in Fig. 2(b). The edge length $a$ varies from 50 nm to 170 nm. The transmission fields indicate that single meta-FP units can also generate high quality colors from purple to red, which shows the ultrahigh resolution of 141111 DPI.

With such high resolution, it is possible to realize dense color mixing. The mixing pixels [Fig. 2(c)], comprising 2×2 meta-FP units, are simulated and analyzed by respectively varying the edge lengths $a_1$ and $a_2$ of the two nano-cuboid groups marked with blue and red. We enlarge the size scope of $a$ as 40 nm to 160 nm to cover more colors. As depicted in Fig. 2(c), colors from the two mixing types cover many of the area within the 194% sRGB gamut. We further print a colorful picture with varying hue and saturation by 30×30 Type 1 mixing pixels, and check its color distribution through simulation. During the calculation (Supplement 3), the numerical aperture is reduced to NA = 1 corresponding to the resolution limit in free space. Fig. 2(d, e) shows the geometry and the simulated color distribution (without brightness information) of the printing. The clear and colorful image proves that paintings and filters using the 2×2 mixing pixels are practical. Moreover, details of single meta-FP units are nearly invisible in free space, which reveals the excellent combination of the multiple color generation ability and the diffraction limit resolution.

For most applications, light usually obliquely exposures onto the printed patterns, therefore requiring the colors to stay unchanged under different incident angles. Fortunately, the high refractive index of Si benefits the FP cavity with enhanced view angles [17, 24], which also works in our design. To verify this, we set up a group of comparison between Si meta-FP cavities and PDMS FP cavities with similar peak wavelengths, simulating their output spectrums under different incident angles widely ranging from 0° to 65° [Fig. 3(a, b)]. Overall, for the meta-FP cavity, the mean relative deviation of the peak wavelength reaches 1%, 2%, 3%, 4% when the incident angle enlarges to 23.6°, 35.4°, 44.4°, 55.9°, and maximal 4.5% at 65°. At the same angles, the mean relative deviations of the PDMS FP cavities are as large as 2.8%, 5.9%, 8.7%, 12.4%, 15.1%, respectively 280%, 316%, 290%, 310%, 335% greater than those of the meta-FP cavities. This indicates that the embedded high refractive index metasurface compresses the angular induced transmission peak deviation by more than 300%. To directly reflect such improvement on the structural color, we calculate output colors from both meta-FP cavities and PDMS FP-cavities at different oblique angles (Fig. 3(c)). The chromatism is clearly reduced for meta-FP cavities.

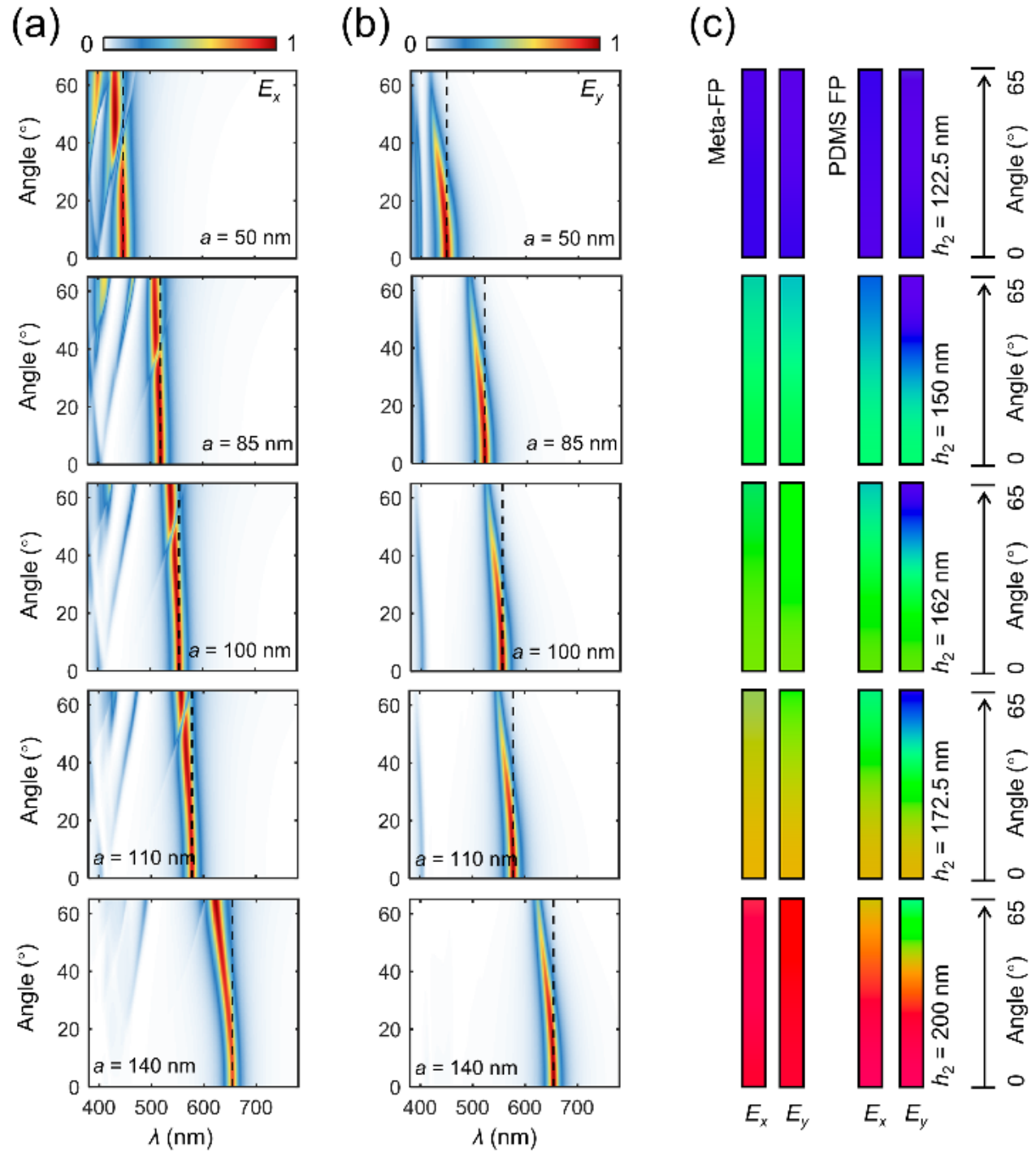

**Fig. 3.** Angular responses of meta-FP cavities and PDMS FP cavities. (a) Simulated transmission spectrums of meta-FP cavities with selected edge lengths $a$ = 50 nm, 85 nm, 100 nm, 110 nm, 140 nm, for $x$ (upper) and $y$ (lower) linearly polarized light, and for oblique incidence from 0° to 65° in the $xz$ plane. (b) Transmission spectrums of PDMS FP cavities with selected PDMS heights $h_2$ = 122.5 nm, 150 nm, 162 nm, 172.5 nm, 200 nm, under the same illumination conditions. (c) Calculated colors of above meta-FP cavities and PDMS FP cavities. Colors from meta-FP cavities show significantly reduced dependence on incident angles.

Furthermore, the embedded metasurface units can introduce flexible polarization manipulation ability which is one of the best-known advantages of gradient metasurfaces. As shown in Fig. 4(a), instead of the nano-cuboid, the cross shaped nanopillar with arm lengths $w_1$, $w_2$ and fixed arm width $w_0$ = 42 nm is embedded into the PDMS spacer to obtain the polarization-sensitive optical response. The nanopillar height $h_1$ and cavity length $h_2$ are adjusted to 70 nm and 100 nm to increase the color tuning range. Under $x$ linearly polarized normal incidence, we sweep arm lengths $w_1$ and $w_2$ from 42 nm to 160 nm to get colors generated from the meta-FP units. For $y$ linear polarization, because of the geometrical symmetry, colors versus arm lengths are obtained by transposing the former results. Fig. 4(b) shows high purity transmission colors and their corresponding arm lengths $w_1$ and $w_2$.

To show the polarization multiplexing ability, we first print a simple sample by 60×60 units, which separately displays "META" or "NANO" when illuminated by $x$ or $y$ polarized light. Totally 5 colors are used and circles in Fig. 4(b) mark all the meta-FP units in need. Fig. 4(c, d) respectively depict the arm length maps, and the generated pictures without brightness information. Next, two complex printings, displaying the pseudo color images of a ball and a cube, are separately encoded into $x$ and $y$ polarizations with 60×60 units. The geometries of nanopillars [Fig. 4(f)] are chosen following the color requirements at every position. The smallest pixel is set to 1×1 meta-FP unit to take use of the ultrahigh resolution of 141111 DPI. As shown in Fig. 4(g), two paintings are both well generated without any overlap or interference between each other. These two samples indicate that the designed meta-FP cavity enables arbitrary pure color arrangements between cyan and red for orthogonal linear polarization states, also under the sub diffraction limit resolution. By unlocking more adjustable geometrical parameters like two arm widths, the wider color switching is promising. Moreover, such design is also compatible with the color mixing pixels discussed in the previous part, which brings extreme density and flexibility to colorful multiplexed optical storage and encryption.

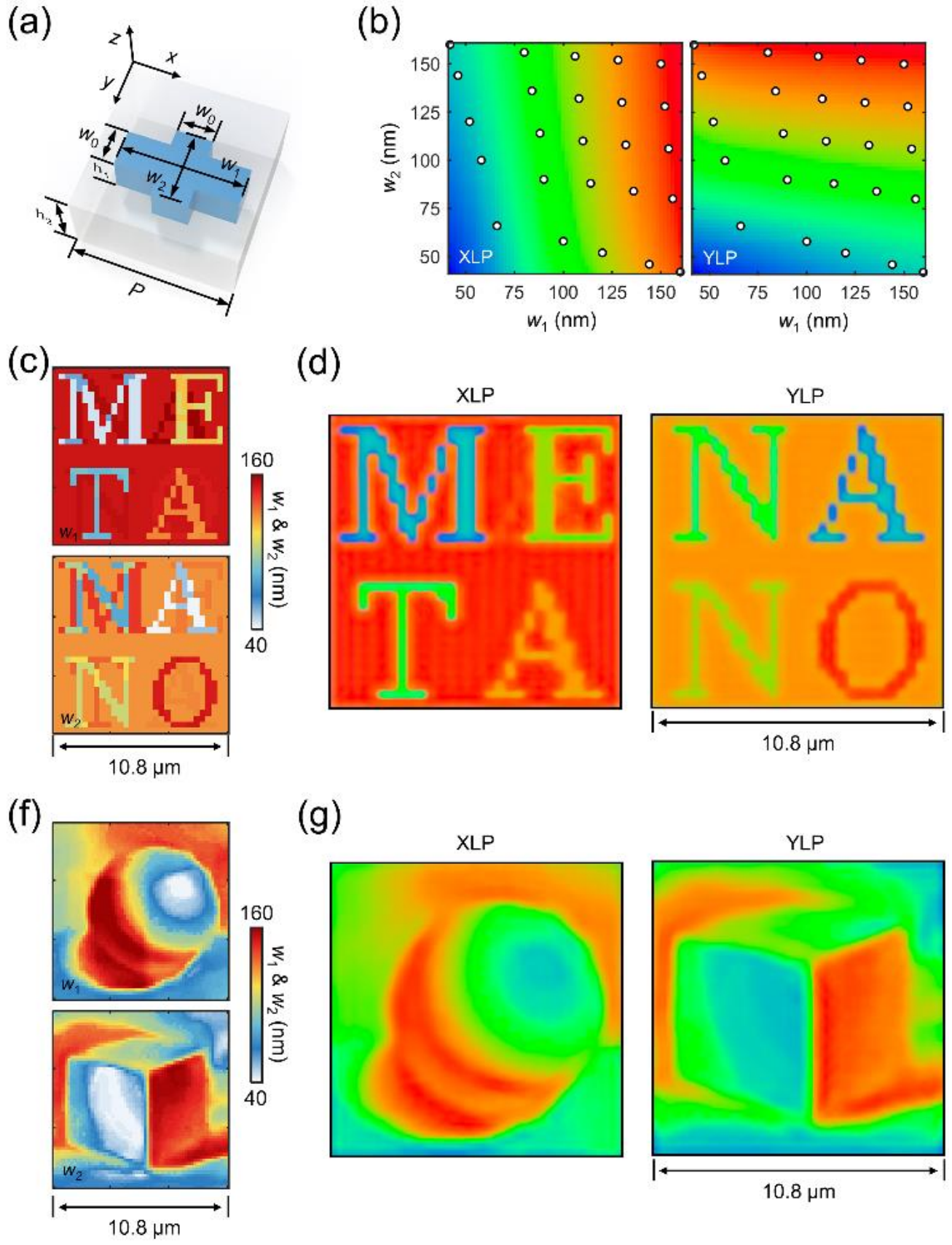

**Fig. 4.** Polarization multiplexed display with the meta-FP cavity. (a) Scheme of a cross shaped metasurface unit. (b) Colors from meta-FP cavities, for $x$ (left) and $y$ (right) linear polarized normal incidence. Circles indicate all unit geometries required for the following printing. (c) Maps of arm lengths $w_1$ (upper) and $w_2$ (lower). All of the used units are marked in (b). (d) Corresponding simulated color distributions under $x$ (left) and $y$ (right) linearly polarized normal illuminations. The brightness information is removed to clearly show the hue. (f, g) Geometries and corresponding simulated images of the second sample.

In summary, we have proposed and simulated Si metasurface embedded Fabry-Pérot cavities to generate high quality transmission structural colors with ultrawide gamut of 194% sRGB, ultrahigh resolution of 141111 DPI and 300% enhanced angular tolerance. The meta-FP cavities also decouple the colors of orthogonal linear polarization states and enable multiplexed encodings. Our meta-FP cavities can be applied in most occasions in need of transmission structural colors, like CMOS and display filters, optical encryption and information storage. As a connection between metasurfaces and colors, further studies might be able to introduce more advantages of gradient metasurfaces into cavities and structural colors.

**Funding.** This work is supported by the Natural Science Foundation of China (No.62075073, 62135004 and 62075129), the Fundamental Research Funds for the Central Universities (No. 2019kfyXKJC038), State Key Laboratory of Advanced Optical Communication Systems and Networks, Shanghai Jiao Tong University (No. 2021GZKF007), and Key R & D project of Hubei Province (No. 2021BAA003).

**Disclosures.** The authors declare no conflicts of interest.

**Data availability.** Data underlying the results presented in this paper are not publicly available at this time but may be obtained from the authors upon reasonable request.

**Supplemental document**. See Supplement 1 for supporting content.

## References

1. T. Lee, J. Jang, H. Jeong, and J. Rho, Nano Converg. 5, 1 (2018).
2. S. Daqiqeh Rezaei, Z. Dong, J. You En Chan, J. Trisno, R. J. H. Ng, Q. Ruan, C.-W. Qiu, N. A. Mortensen, and J. K. W. Yang, ACS Photonics 8, 18-33 (2020).
3. Z. Dong, J. Ho, Y. F. Yu, Y. H. Fu, R. Paniagua-Dominguez, S. Wang, A. I. Kuznetsov, and J. K. W. Yang, Nano Lett. 17, 7620-7628 (2017).
4. I. Koirala, S. S. Lee, and D. Y. Choi, Opt. Express 26, 18320-18330 (2018).
5. B. Yang, W. Liu, Z. Li, H. Cheng, S. Chen, and J. Tian, Adv. Opt. Mater. 6, 1701009 (2018).
6. W. Yang, S. Xiao, Q. Song, Y. Liu, Y. Wu, S. Wang, J. Yu, J. Han, and D. P. Tsai, Nat. Commun. 11, 1864 (2020).
7. E. Balaur, C. Sadatnajafi, S. S. Kou, J. Lin, and B. Abbey, Sci. Rep. 6, 28062 (2016).
8. X. Duan, S. Kamin, and N. Liu, Nat. Commun. 8, 14606 (2017).
9. N. Pinton, J. Grant, S. Collins, and D. R. S. Cumming, ACS Photonics 5, 1250-1261 (2018).
10. Y. Jung, H. Jung, H. Choi, and H. Lee, Nano Lett. 20, 6344-6350 (2020).
11. Z. Yang, Y. Chen, Y. Zhou, Y. Wang, P. Dai, X. Zhu, and H. Duan, Adv. Opt. Mater. 5, 1700029 (2017).
12. Y. Wang, M. Zheng, Q. Ruan, Y. Zhou, Y. Chen, P. Dai, Z. Yang, Z. Lin, Y. Long, Y. Li, N. Liu, C. W. Qiu, J. K. W. Yang, and H. Duan, Research (Wash D C) 2018, 8109054 (2018).
13. A. M. Shaltout, J. Kim, A. Boltasseva, V. M. Shalaev, and A. V. Kildishev, Nat. Commun. 9, 2673 (2018).
14. L. Pjotr Stoevelaar, J. Berzins, F. Silvestri, S. Fasold, K. Zangeneh Kamali, H. Knopf, F. Eilenberger, F. Setzpfandt, T. Pertsch, S. M. B. Baumer, and G. Gerini, Opt. Express 28, 19818-19836 (2020).
15. A. McClung, S. Samudrala, M. Torfeh, M. Mansouree, and A. Arbabi, Science Advances 6 (2020).
16. X. Zhu, W. Yan, U. Levy, N. A. Mortensen, and A. Kristensen, Science Advances 3 (2017).
17. J. Zhao, M. Qiu, X. Yu, X. Yang, W. Jin, D. Lei, and Y. Yu, Adv. Opt. Mater. 7, 1900646 (2019).
18. L. Li, J. Niu, X. Shang, S. Chen, C. Lu, Y. Zhang, and L. Shi, ACS Appl. Mater. Interfaces 13, 4364-4373 (2021).
19. J. S. Lee, J. Y. Park, Y. H. Kim, S. Jeon, O. Ouellette, E. H. Sargent, D. H. Kim, and J. K. Hyun, Nat. Commun. 10, 4782 (2019).
20. I. H. Lee, G. Li, B. Y. Lee, S. U. Kim, B. Lee, S. H. Oh, and S. D. Lee, Opt. Express 27, 24512-24523 (2019).
21. D. P. Edward, and I. Palik, "Handbook of optical constants of solids," (Academic, Orlando, Florida, 1985).
22. V. Gupta, P. T. Probst, F. R. Gossler, A. M. Steiner, J. Schubert, Y. Brasse, T. A. F. Konig, and A. Fery, ACS Appl. Mater. Interfaces 11, 28189-28196 (2019).

23. P. B. Johnson, and R. W. Christy, Phys. Rev. B 6, 4370-4379 (1972).
24. M. ElKabbash, E. Ilker, T. Letsou, N. Hoffman, A. Yaney, M. Hinczewski, and G. Strangi, Opt. Lett. 42, 3598-3601 (2017).